\documentclass[english]{msjproc}

\usepackage{amssymb,amsmath,amscd}
\usepackage{amsfonts}
\usepackage{young,colortbl}
\usepackage{graphicx}

\usepackage{epstopdf}
\usepackage{graphics}
\newtheorem{Theorem}{Theorem}[section]
\newtheorem{Proposition}{Proposition}[section]

\def\proof{\par{\it Proof}. \ignorespaces}
\def\endproof{{\ \vbox{\hrule\hbox{%
     \vrule height1.3ex\hskip0.8ex\vrule}\hrule }}\par}

\newtheorem{Definition}[Theorem]{Definition}
\newtheorem{Example}[Theorem]{Example}

\newtheorem{Remark}[Theorem]{Remark}
\numberwithin{equation}{section}

\numberwithin{figure}{section}
\def\sech{\mathop{\rm sech}\nolimits}

\begin{document}
  \title{KP Solitons and Mach Reflection in Shallow Water}
  
  \author{Yuji Kodama}{Ohio State University}
  \email{kodama@math.ohio-state.edu}



  \thanks{This work was partially supported by NSF grant DMS-1108813.}

  \maketitle

 \begin{abstract}
 This talk gives a survey of our recent studies on soliton solutions of the  Kadomtsev-Petviashvili equation with an emphasis on the Mach reflection problem in shallow water.
     \end{abstract}

\section{Shallow water waves: Basic equations}
   We consider a surface wave
on water which is assumed to be irrotational and incompressible  (see for examples \cite{Wh:74, AS:81}).
We are mainly interested in a long wave phenomena, and assume zero surface tension, that is, 
we ignore the gravity-capillary waves.  Let us denote the following scales:
\begin{align*}
&\lambda_0~\sim~{\rm horizontal~length~scale=typical~wave~length}\\
&h_0~\sim~{\rm vertical~length~scale=asymptotic~water~depth}\\
&a_0~\sim~{\rm nonlinear~scale=typical~wave~amplitude}
\end{align*}
The non-dimensional variables $\{x,y,z,t,\eta,\phi\}$ with the corresponding
physical variables with  the space variables $(\tilde{x},\tilde{y},\tilde{z})$, 
the amplitude of the wave $\tilde\eta$, and the velocity potential $\tilde\phi$ are defined as
\begin{equation}\label{Pvariables}
\left\{\begin{array}{llll}
\displaystyle{\tilde x=\lambda_0 x,\quad \tilde y=\lambda_0 y,\quad \tilde z=h_0 z,\quad \tilde t=\frac{\lambda_0}{c_0}t},\\[1.5ex]
\displaystyle{\tilde\eta=a_0\eta,\qquad \tilde\phi=\frac{a_0}{h_0}\lambda_0c_0\phi.}
\end{array}\right.
\end{equation}
Then the shallow water equation in the non-dimesional form is given by
\[
\begin{array}{llll}
~~\phi_{zz}+\beta\Delta\phi=0,   \qquad &{\rm for}\quad  0<  z <1+\alpha\eta, \\[1.0ex]
~~\phi_{ z}=0, \qquad &{\rm at}\quad z=0,\\[1.0ex]
\left.\begin{array}{llll}
\displaystyle{\phi_{ t}+\frac{1}{2}\alpha\left|\nabla\phi\right|^2+\frac{1}{2}\frac{\alpha}{\beta}\phi^2_z+\eta=0,} \\[2.0ex]
\displaystyle{\eta_{t}+\alpha\nabla\phi\cdot\nabla\eta=\frac{1}{\beta}\phi_{ z}},
\end{array}\right\}\quad&{\rm at}\quad z=1+\alpha \eta,\\
\end{array}
\]
where $\nabla=(\frac{\partial}{\partial x}, \frac{\partial}{\partial y})$ and $\Delta=\nabla^2$,
the two-dimensional Laplace operator, for $(x,y)\in\mathbb{R}^2$.
The parameters $\alpha$ and $\beta$ are given by
\[
\alpha=\frac{a_0}{h_0}\qquad {\rm and}\qquad \beta=\left(\frac{h_0}{\lambda_0}\right)^2.
\]
The weak nonlinearity implies $\alpha\ll 1$, and the weak dispersion (or long wave assumption)
implies $\beta\ll 1$. With a small parameter $\epsilon\ll 1$, we assume $
\alpha\sim \beta =\mathcal O(\epsilon)$.

First we note, from the first two equations, that $\phi$ can be written formally in the form,
\[
\phi(x,y,z,t)=\cos\left(z\sqrt{\beta\Delta}\right)\psi=\psi-\beta\frac{z^2}{2}\Delta\psi +\beta^2\frac{z^4}{24}\Delta^2\psi+\mathcal O(\epsilon^3).
\]
where $\psi(x,y,t)=\phi(x,y,0,t)$.
Then the equations at the surface $z=1+\alpha\eta$ give a Boussinesq-type system,
\begin{align}\nonumber
\psi_t+\eta+&\frac{\alpha}{2}\left|\nabla\psi\right|^2-\frac{\beta}{2}\Delta\psi_t\\
+&\frac{\alpha\beta}{2}\left(\left(\Delta\psi\right)^2+\nabla\psi\cdot\nabla(\Delta\psi)-2\eta\Delta\psi_t\right)+\frac{\beta^2}{24}\Delta^2\psi_t=\mathcal{O}(\epsilon^3).
\label{eq:eta}\\
\eta_t+\Delta\psi+&\alpha\nabla\cdot\left(\eta\nabla\psi\right)-\beta\frac{1}{6}\Delta^2\psi\nonumber \\
-&\frac{\alpha\beta}{2}\left(\nabla\eta\cdot\nabla(\Delta\psi)+\eta\Delta^2\psi\right)+\frac{\beta^2}{120}\Delta^3\psi=\mathcal O(\epsilon^3).
\label{eq:etat}
\end{align}

We now derive the KP equation with higher order corrections which will be a key equation
for a physical application.  The KP equation is obtained 
under the assumption with a weak dependence in the $y$-direction (quasi-two-dimensionality), and we
introduce a small parameter $\gamma$ so that the $y$-coordinate is scaled as
\begin{equation}\label{eq:zeta}
\zeta :=\sqrt{\gamma} y,  \qquad {\rm with}\quad \gamma=\mathcal O(\epsilon).
\end{equation}
 We also consider a far field with the scaled coordinates 
$\xi=x-t$ and $\tau=\epsilon t$ (i.e. a unidirectional approximation).
Then eliminating $\eta$ in the equations \eqref{eq:eta} and \eqref{eq:etat},  we obtain the KP equation for the function $v:=\psi_{\xi}(\xi,\zeta,\tau)$ with higher order corrections up to $\mathcal{O}(\epsilon^2)$,
\begin{align}\nonumber
2\epsilon v_{\tau}+&3\alpha vv_{\xi}+\frac{\beta}{3}v_{\xi\xi\xi}+\gamma D^{-1}v_{\zeta\zeta} \\
&+\frac{19}{180}\beta^2v_{\xi\xi\xi\xi\xi}+\alpha\beta\left(\frac{15}{6}v v_{\xi\xi\xi}+\frac{53}{12}v_{\xi}v_{\xi\xi}\right)+\frac{\beta\gamma}{2}v_{\xi\zeta\zeta}-\frac{\gamma^2}{4}D^{-3}v_{\zeta\zeta\zeta\zeta}
 \nonumber\\
&+\alpha\gamma\left(\frac{5}{4}v
D^{-1}v_{\zeta\zeta}+2v_{\zeta}D^{-1}v_{\zeta}-\frac{3}{4}D^{-1}(v^2)_{\zeta\zeta}+\frac{1}{2}v_\xi
D^{-2}v_{\zeta\zeta}\right)=\mathcal{O}(\epsilon^3).\label{hKP}
\end{align}
where $D^{-1}:=\partial_{\xi}^{-1}$ is a formal integral operator.
The first line of this equation of the order $\mathcal{O}(\epsilon)$ is the KP equation, and the terms
in the second and third lines are the $\epsilon^2$-order corrections to the KP equation.
The wave amplitude $\eta$ is then given by
\begin{equation}\label{eta}
\eta=v+\frac{\alpha}{4}v^2-\frac{\beta}{3}v_{\xi\xi}+\frac{\gamma}{2}D^{-2}v_{\zeta\zeta}+\mathcal{O}(\epsilon^2).
\end{equation}
\begin{Remark}
In terms of physical coordinates, the KP equation is given by
\[
\left(\tilde{\eta}_{\tilde t}+c_0\tilde{\eta}_{\tilde x}+\frac{3c_0}{2h_0}\tilde{\eta}\tilde{\eta}_{\tilde x}+
\frac{c_0h_0^2}{6}\tilde{\eta}_{\tilde x\tilde x\tilde x}\right)_{\tilde x}+\frac{c_0}{2}\tilde{\eta}_{\tilde y\tilde y}=0.
\]
As a particular solution, we have a solitary wave solution in the coordinate perpendicular
to the wave crest, $\tilde\chi=\tilde{x}\cos\Psi_0+\tilde{y}\sin\Psi_0$,
\begin{equation}\label{KPsoliton}
\tilde \eta=a_0{\rm sech}^2\sqrt{\frac{3a_0}{4h_0^3\cos^2\Psi_0}}\left[\,\tilde\chi-c_0\cos\Psi_0\left(1+\frac{a_0}{2h_0}
+\frac{1}{2}\tan^2\Psi_0\right)\,\tilde t -\chi_0\,\right].
\end{equation}
where $a_0>0, \Psi_0$ and $\tilde x_0$ are arbitrary constants.
Recall that the KP equation is derived under the assumption of quasi-two dimensionality,
that is, $\gamma=\tan^2\Psi_0=\mathcal{O}(\epsilon)$, and the solution \eqref{KPsoliton} becomes unphysical for the case with a large angle. 
That is, the width of the solitary wave depends on the propagation direction, $\Psi_0$,
which should be corrected when we apply the KP equation to a physical problem.
In this talk,  I will explain  how this can be fixed using a {\it normal form} theory
concerning the higher order corrections to the KP equation, and
show that some of the exact solutions of the KP equation can be used to describe
two-dimensional interaction phenomena in shallow water, called the Mach reflection.
\end{Remark}


\section{Normal form for the KP equation with higher order corrections}
In order to study the higher order corrections, 
 we define a {\it normal form} for \eqref{hKP}. For this purpose, we first put \eqref{hKP} into
a canonical form of the KP equation with the change of variables,
\[
\xi=\sqrt{\beta}\,X,\qquad \zeta=\sqrt{\beta\gamma}\,Y,\qquad \tau=-\frac{3\epsilon\sqrt{\beta}}{2}\,T,\qquad
v=\frac{2}{3\alpha}u.
\]
Then \eqref{hKP} becomes
\begin{align}\nonumber
4u_{T}&=6uu_{X}+u_{XXX}+3 D^{-1}u_{YY} \\
&+\frac{19}{60}u_{XXXXX}+\frac{5}{3}uu_{XXX}+\frac{53}{6}u_{X}u_{XX}+\frac{3}{2}u_{XYY}-\frac{3}{4}D^{-3}u_{YYYY} \nonumber\\
&+\frac{5}{2}uD^{-1}u_{YY}+4u_{Y}D^{-1}u_{Y}-\frac{3}{4}D^{-1}(u^2)_{YY}+u_X
D^{-2}u_{YY}+\mathcal{O}(\epsilon^{\frac{9}{2}}).\label{hKP2}
\end{align}
Note here we have the orders $u\sim\mathcal{O}(\epsilon)$, $\partial_X\sim \mathcal{O}(\epsilon^{\frac{1}{2}})$, $D^{-1}\sim\mathcal{O}(\epsilon^{-\frac{1}{2}})$, and $\partial_Y\sim\mathcal{O}(\epsilon)$.  Here the new variables $(X,Y,T)$ are related to the physical ones with
\begin{equation}\label{realvariables}
\tilde x-c_0\tilde t=h_0X\,,\qquad \tilde y=h_0\,Y,\qquad \tilde t=\frac{3h_0}{2c_0}\,T.
\end{equation}
The wave amplitude $\eta$ in terms of $u$  is  given by
\[
\alpha\eta=\frac{2}{3}u+\frac{1}{9}u^2-\frac{2}{9}u_{XX}+\frac{1}{3}D^{-2}u_{YY}+\mathcal{O}(\epsilon^3)
\]

Hereafter we use the lower case letters $(x,y,t)$ for $(X,Y,T)$, and the KP variables can be converted to the physical
variables directly through the relations \eqref{realvariables}.

\subsection{The normal form for the KdV equation}
Before discussing a normal form for \eqref{hKP}, we give a brief summary of the result in \cite{K:85,K:87,HK:09}
for the case of the KdV equation.  Taking $\partial_Yu=0$ in \eqref{hKP2},
we have the KdV equation with higher order corrections,
\begin{equation}\label{hkdv}
4u_{t}=6uu_{x}+u_{xxx}+
\left(\frac{19}{60}u_{xxxxx}+\frac{5}{3}uu_{xxx}
 +\frac{53}{6}u_{x}u_{xx}\right)+\mathcal{O}(\epsilon^{\frac{9}{2}}).
\end{equation}
In \cite{K:85,K:87,HK:09}, we found that one can transform \eqref{hkdv} with a formal change of variable, a Lie exponential transformation $e^{V_{\varphi}}$ with the generating function $\varphi$  (see \cite{HK:09}),
\begin{equation}\label{LieKdV}
u=e^{V_{\varphi}}\circ U=U+\left(U_{xx}+\frac{4}{3}U^2+\frac{1}{2}U_xD^{-1}U\right)+\mathcal{O}(\epsilon^3),
\end{equation}
into the equation which we referred to as the {\it normal form} of \eqref{hkdv},
\[
4U_{t}=6UU_{x}+U_{xxx}+
\frac{19}{60}\left(U_{xxxxx}
 +10UU_{xxx}+20U_xU_{xx}+30U^2U_x\right)+\mathcal{O}(\epsilon^{\frac{9}{2}}).
 \]
The point here is that the higher order term of  $\mathcal{O}(\epsilon^{\frac{7}{2}})$
in this equation is the 5th order symmetry of the
KdV equation, and hence the normal form is integrable up to $\mathcal{O}(\epsilon^{\frac{7}{2}})$.
That is, we have an integrability  not only at the KdV of $\mathcal{O}(\epsilon^{\frac{5}{2}})$
but also at the next order correction of  $\mathcal{O}(\epsilon^{\frac{7}{2}})$.
This is true even for the general form of the higher order correction,
\[
\alpha_1u_{xxxxx}+\alpha_2uu_{xxx}+\alpha_3u_xu_{xx}+\alpha_4u^2u_x.
\]
with arbitrary coefficients $\alpha_1,\ldots, \alpha_4$.
This implies that any weakly nonlinear long-wave equation whose
leading order is approximated by  the KdV equation is {\it asymptotically integrable} 
up to the next order approximation \cite{K:85,K:87,HK:09}.

Note also that the normal form admits one-soliton solution in the form,
\[
U=A_0\sech^2\sqrt{\frac{A_0}{2}}(x+x_0(t)),
\]
where $x_0(t)$ is determined by 
$
\frac{dx_0}{dt}:=C_{\rm hKdV}=\frac{1}{2}A_0+\frac{19}{60}A_0^2+\mathcal{O}(\epsilon^3).
$
Then the solution of the higher order KdV equation \eqref{hkdv} is given by the transformation
\eqref{LieKdV}, i.e.
\[
u=A_0S^2+\left(A_0^2S^2-\frac{2}{3}A_0^2S^4\right)+\mathcal{O}(\epsilon^3).
\]
where $S:=\sech\sqrt{\frac{A_0}{2}}(x+x_0)$.  Notice that the amplitude $\eta$ of \eqref{eta} for the KdV case
is given by
\begin{align}\label{KdVamp}
\alpha\eta=\frac{2}{3}A_0S^2+\frac{2}{9}A_0^2S^2+\frac{1}{3}A_0^2S^4+\mathcal{O}(\epsilon^3)
\end{align}

\subsection{Normal form of the KP equation}
Now we define a normal form of \eqref{hKP2} by a Lie exponential transform,
\[
u=e^{V_{\varphi}}\circ U=U+\left(\beta_1 U_{xx}+\beta_2U^2+\beta_3U_xD^{-1}U+\beta_4D^{-2}U_{yy}\right)+\mathcal{O}(\epsilon^3),
\]
where $\beta_j$'s are determined such a way that the transformed equation (normal form) 
has a {``good''} property.  First, we require that the normal form is reduced to the KdV-normal form
when we have $\partial_yu=0$.  That is, from \eqref{LieKdV}, we have
$
\beta_1=1, \beta_2=\frac{4}{3}, \beta_3=\frac{1}{2},
$
but $\beta_4$ remains free.  Then we require that the normal form admits a solitary wave
in the form of one-soliton solution, 
\[
U=A_0\sech^2\sqrt{\frac{A_0}{2}}(x+x_0(y,t)).
\]
This determines uniquely $\beta_4=\frac{1}{2}$, and the normal form of \eqref{hKP2} is given by
\begin{align}\nonumber
4U_{t}&=6UU_{x}+U_{xxx}+3 D^{-1}U_{yy} \\
&+\frac{19}{60}\left(U_{xxxxx}+10UU_{xxx}+20U_{x}U_{xx}+30U^2U_x\right)+\frac{3}{2}U_{xyy}-\frac{3}{4}D^{-3}U_{yyyy}  \nonumber\\
&-UD^{-1}U_{yy}+7U_{y}D^{-1}U_{y}+\frac{1}{4}D^{-1}(U^2)_{yy}+\frac{5}{2}U_x
D^{-2}U_{yy}+\mathcal{O}(\epsilon^{\frac{9}{2}}).\label{N-KP2}
\end{align}
Then the phase $x_0(y,t)$ satisfies
\[
4\frac{\partial x_0}{\partial t}=2A_0+3\left(\frac{\partial x_0}{\partial y}\right)^2+\frac{19}{15}A_0^2+3A_0\left(\frac{\partial x_0}{\partial y}\right)^2-\frac{3}{4}\left(\frac{\partial x_0}{\partial y}\right)^4 +\mathcal{O}(\epsilon^3).
\]
Setting the angle $\Psi_0$ of the soliton with $\frac{\partial x_0}{\partial y}=\tan\Psi_0$, we have the velocity $C_{\rm hKP}=\frac{\partial x_0}{\partial t}$
of the KP soliton with the higher order corrections.
The corresponding solitary wave solution $u$ is then given by
\[
u=A_0S^2+\left(A_0^2S^2+\frac{1}{2}A_0\tan^2\Psi_0S^2-\frac{2}{3}A_0^2S^4\right)+\mathcal{O}(\epsilon^3)
\]
and the wave amplitude $\eta$ of \eqref{eta} is given by
\begin{align}\nonumber
\alpha\eta&=\frac{2}{3}A_0S^2+\frac{2}{3}\tan^2\Psi_0S^2+\frac{2}{9}A_0^2S^2+\frac{1}{3}A_0^2S^4+\mathcal{O}(\epsilon^3)\\
&=\frac{2}{3}[A_0]S^2+\frac{2}{9}[A_0]^2S^2+\frac{1}{3}[A_0]^2S^4+\mathcal{O}(\epsilon^3),
\label{KPamp}
\end{align}
where $[A_0]:=A_0(1+\tan^2\Psi_0)={A_0}/{\cos^2\Psi_0}$ (cf.\eqref{KPsoliton}). Notice that the correction to quasi-two-dimensional approximation can be absorbed into the amplitude of the KdV equation (cf.\eqref{KdVamp}). This formula gives the relation between the observed amplitude
$\eta$ from numerical simulation (or experiment)  of shallow water wave system and
the KP amplitude $A_0$.  


\section{The KP solitons}
Here we give a brief summary of the soliton solutions of the KP equation \cite{CK:09,K:04,K:10},
\begin{equation}\label{KPS}
-4u_t+6uu_x+u_{xxx}+D^{-1}u_{yy}=0.
\end{equation}
We write the solution of the KP equation in the $\tau$-function form,
\begin{equation}\label{utau}
u(x,y,t)=2\partial_x^2\ln\tau(x,y,t).
\end{equation}
where the $\tau$-function is assumed to be the Wronskian determinant with $N$ functions
$f_i$'s (see for examples \cite{Sa:79,H:04}),
\begin{equation}\label{tau}
\tau={\rm Wr}(f_1,f_2,\ldots,f_N).
\end{equation}
Here the
functions $\{f_1,\ldots,f_N\}$ form a set of linearly independent
solutions of the linear equations,
\[
{\partial_y}f_n={\partial_x^2}f_n,\qquad {\partial_t}f_n={\partial_x^3}f_n.
\]
In particular, we
consider a set of finite dimensional solutions for $\{f_1,f_2,\ldots,f_N\}$,
\[
f_n(x,y,t)=\sum_{m=1}^Ma_{n,m}E_m(x,y,t)\qquad {\rm with}\quad E_m=e^{\theta_m}:=\exp({\kappa}_mx+{\kappa}_m^2y+{\kappa}_m^3t).
\]
Thus this type of solution is characterized by the {\it ordered} parameters $\{{\kappa}_1<\cdots<{\kappa}_M\}$ and
the $N\times M$ matrix $A:=(a_{n,m})$ of rank$(A)=N$, that is, we have
\begin{equation}\label{NM}
(f_1,f_2,\ldots,f_N)=(E_1,E_2,\ldots, E_M)A^T\,.
\end{equation}
Note that $\{E_1,E_2,\ldots,E_M\}$ gives a basis of $\mathbb{R}^M$ and $\{f_1,f_2,\ldots,f_N\}$
spans an $N$-dimensional subspace of $\mathbb{R}^M$. This means that the $A$-matrix
can be identified as a point on the real Grassmann manifold Gr$(N,M)$ (see \cite{K:04,CK:09}).
More precisely, let $M_{N\times M}({\mathbb R})$ be the set of all $N\times M$ matrices of rank $N$.
Then Gr$(N,M)$ can be expressed as
\[
{\rm Gr}(N,M)={\rm GL}_N({\mathbb R})\backslash M_{N\times M}({\mathbb R}).
\]
where GL$_N(\mathbb R)$ is the general linear group of rank $N$.  This expression means that
other basis $(g_1,\ldots,g_N)=(f_1,\ldots,f_N)H$ for any $H\in{\rm GL}_N({\mathbb R})$
spans the same subspace, that is, $A \to H^TA$ (GL$_N(\mathbb R)$ acting from the left).
 Notice here that the freedom in the $A$-matrix with GL$_N({\mathbb R})$ can be fixed by expressing $A$ in the reduced row echelon form (RREF).

Now using the Binet-Cauchy Lemma for the determinant, the $\tau$-function of (\ref{tau})
can be expressed in the form,
\begin{align}
\tau(x,y,t)&=\sum_{J\in\binom{[M]}{N}}\Delta_J(A)E_J(x,y,t), \label{B-C}
\end{align}
where $\Delta_J(A)$ is the $N\times N$ minor of the $A$-matrix with $N$ columns
marked by $J=\{j_1,\ldots,j_N\}\in\binom{[M]}{N}$, the set of $N$-subindices of the set $[M]:=\{1,\ldots, M\}$, and $E_J(x,y,t)$ is given by
\[
E_J(x,y,t)={\rm Wr}(E_{j_1},\ldots,E_{j_N})=\prod_{l<m}({\kappa}_{j_m}-{\kappa}_{j_l})E_{j_1}\cdots E_{j_N}.
\]
We are also interested in non-singular solutions. Since the solution is given by $u=2\partial_x^2(\ln\tau)$,
the non-singular solutions are obtained by imposing the non-negativity condition on the minors,
\begin{equation}\label{NNC}
\Delta_J(A)\ge 0,\qquad {\rm for~all}\quad J\in\binom{[M]}{N}.
\end{equation}
This condition is  sufficient  for the non-singularity of the solution
for any initial data (see \cite{KW:12} for the necessary condition
for the regularity). We call a matrix $A$ having
the condition (\ref{NNC}) {\it totally non-negative} matrix, referred to as {\it TNN} matrix, and the set of those matrices
forms {\it totally non-negative} Grassmannian, denoted by
Gr$^+(N,M)$ (see e.g. \cite{P:06,KW:11}).

 \begin{Example} Let us express one line-soliton solution  in our setting.  Here we also
 introduce some notations to describe the soliton solutions.
One soliton solution is obtained by the $\tau$-function with $M=2$ and $N=1$, i.e.
 $
 \tau=f_1=a_{11}E_1+a_{12}E_2.
 $
 with $A=(a_{11}~ a_{12})$.
 Since the solution $u$ is given by \eqref{utau}, one can assume $a_{11}=1$ and denote $a_{12}=a>0$.  Then we have
\[
u=2\partial^2_x\ln \tau=\frac{1}{2}({\kappa}_1-{\kappa}_2)^2\sech^2\frac{1}{2}(\theta_1-\theta_2-\ln a).
\]
Thus the solution is localized along the line $\theta_1-\theta_2=\ln a$, hence we call it {\it line-soliton} solution. 
We emphasize here that the line-soliton appears at the boundary of
two regions where either $E_1$ or $E_2$ is the dominant exponential term,
and because of this we also call this soliton a $[1,2]$-soliton solution.
 We refer to each of
these asymptotic line-solitons as the $[i,j]$-soliton.
The $[i,j]$-soliton solution with $i<j$ has the same (local) structure as the
one-soliton solution, and can be described as follows
\begin{equation*}
u=A_{[i,j]}\sech^2\frac{1}{2}\left({\bf K}_{[i,j]}\cdot {\bf x}-\Omega_{[i,j]}t+\Theta^0_{[i,j]}\right)
\end{equation*}
with some constant $\Theta^0_{[i,j]}$. 
The amplitude $A_{[i,j]}$, the wave-vector ${\bf K}_{[i,j]}$ and the frequency $\Omega_{[i,j]}$ are then expressed by
\[
A_{[i,j]}=\frac{1}{2}({\kappa}_j-{\kappa}_i)^2,\qquad
{\bf K}_{[i,j]}=\left({\kappa}_j-{\kappa}_i, {\kappa}_j^2-{\kappa}_i^2\right),\qquad
\Omega_{[i,j]}={\kappa}_j^3-{\kappa}_i^3.
\]
Note that ${\bf K}_{[i,j]}$ and $\Omega_{[i,j]} $ satisfy the soliton-dispersion relation,
$4\Omega_{[i,j]} K_{[i,j]}^x=(K_{[i,j]}^x)^4+3(K_{[i,j]}^y)^2$.
The direction of the wave-vector ${\bf K}_{[i,j]}=(K_{[i,j]}^x,K_{[i,j]}^y)$ is 
measured in the counterclockwise rotation from the $y$-axis, and
it is given by
\[
\frac{K^y_{[i,j]}}{K^x_{[i,j]}}=\tan\Psi_{[i,j]}={\kappa}_i+{\kappa}_j,
\]
that is, $\Psi_{[i,j]}$ gives the angle between the line ${\bf K}_{[i,j]}\cdot {\bf x} =const$ and the $y$-axis.  
\begin{figure}[h]
\begin{centering}
\includegraphics[scale=0.4]{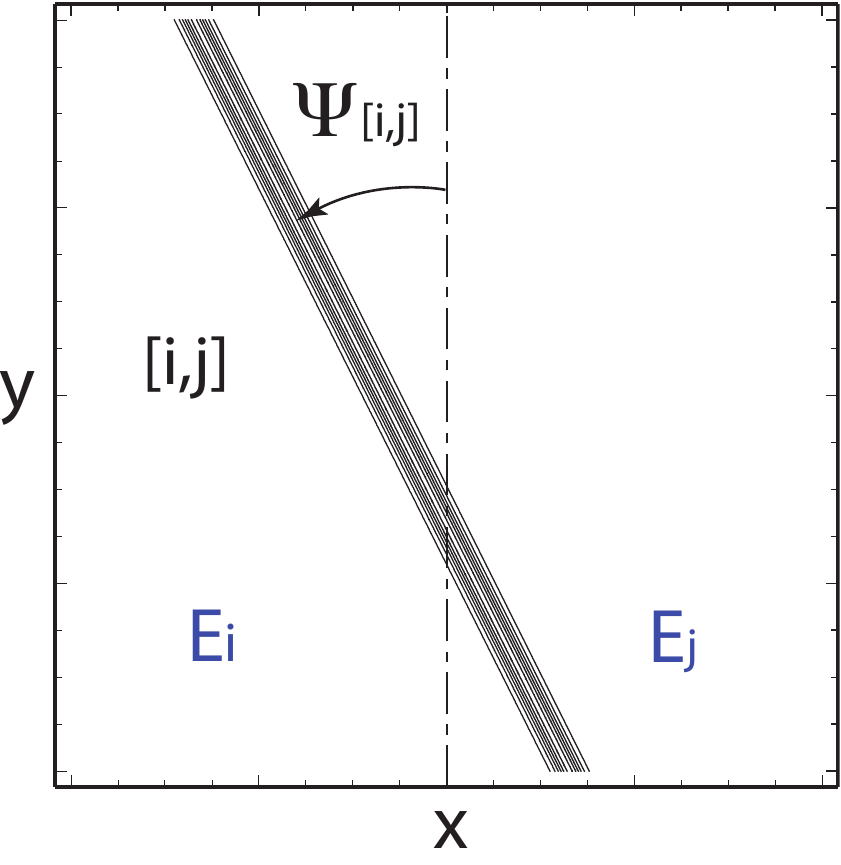}\hskip 3cm
\raisebox{0.3cm}{\includegraphics[height=2.5cm]{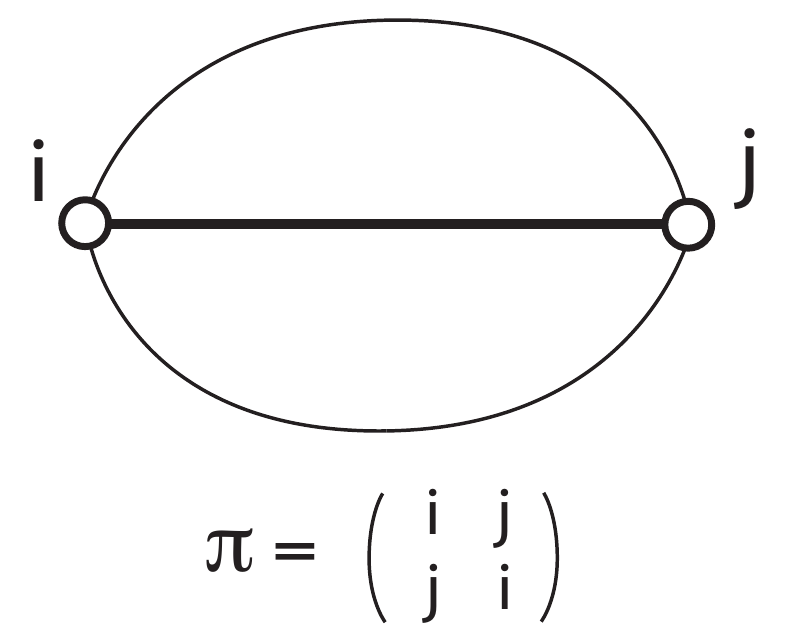}}
\par\end{centering}
\caption{One line-soliton solution of $[i,j]$-type and the corresponding chord diagram. 
The upper oriented chord represents
the part of $[i,j]$-soliton for $y\gg 0$ and the lower one for $y\ll 0$.
\label{fig:1soliton}}
\end{figure}

In  Figure \ref{fig:1soliton}, we illustrate one line-soliton solution of $[i,j]$-type. In the right panel
of this figure, we show a {\it chord diagram} which represents this soliton solution. Here the chord 
diagram indicates the permutation of the dominant exponential terms $E_i$ and $E_j$ 
in the $\tau$-function,  that is, with the ordering
${\kappa}_i<{\kappa}_j$, $E_i$ dominates in $x\ll 0$, while $E_j$ dominates in $x\gg 0$ (see Section \ref{sec:CL}
for the precise definition of the chord diagram).
\end{Example}
 
 \section{Classification of soliton solutions}\label{sec:CL}
We now present the classification theorems of the soliton solutions generated by
the $\tau$-function defined in \eqref{B-C}.
Here we consider the matrix $A$ to be in RREF, and we also assume 
that $A$ is {\it irreducible} as defined below:
\begin{Definition} An $N\times M$ matrix $A$ is {\it irreducible} if each 
column of $A$ contains at least one nonzero element, or each row contains at least
one nonzero element other than the pivot once $A$ is in RREF.
So the irreducibility implies that we consider only derangements (i.e. no fixed points) of the permutation.
\end{Definition}

Our classification scheme of the soliton solutions is given by 
 identifying the asymptotic line-solitons as $y \to\pm \infty$. We denote a
line-soliton solution by $(N_-,N_+)$-soliton whose asymptotic form
consists of $N_-$ line-solitons as $y\to-\infty$ and $N_+$ line-solitons
for $y\to\infty$ in the $xy$-plane.
The next Proposition provides a general result characterizing
the asymptotic line-solitons of the $(N_-,N_+)$-soliton solutions (the proof can be found in \cite{CK:09}):
\begin{Proposition}\label{asymptote}
Let $\{e_1,e_2,\ldots,e_N\}$ and $\{g_1,g_2,\ldots,g_{M-N}\}$ denote respectively,
the pivot and non-pivot indices associated with
an irreducible, $N\times M$, TNN $A$-matrix. Then the soliton 
solution obtained from the $\tau$-function in (\ref{B-C}) with this $A$-matrix has
the following structure:
\begin{itemize}
\item[(a)] For $y\gg 0$, there are $N$  line-solitons of  $[e_n, j_n]$-type for some $j_n$.
\item[(b)] For $y\ll 0$, there are $(M-N)$  line-solitons of $[i_m, g_m]$-type for some $i_m$.
\end{itemize}
\label{engn}
\end{Proposition}

An important consequence of Proposition \ref{engn} is 
that it defines the {\it pairing} map $\pi: [M] \to [M]$ with
\begin{equation}
\left\{\begin{array}{lll}
\pi(e_n) &= j_n\,,\quad & n=1,2,\ldots,N\,, \\[1.0ex]
\pi(g_m) &= i_m\,, \quad &m=1,2,\ldots,M-N\,.
\end{array}\right.
\label{pi}
\end{equation}
Recall that $\{e_n\}_{n=1}^N$ and $\{g_m\}_{m=1}^{M-N}$ are respectively, 
the pivot and non-pivot indices of the $A$-matrix and form a disjoint partition 
of $[M]$. Then the unique index pairings in Proposition \ref{engn} imply that
the map $\pi$ is a {\it permutation} of $M$ indices, i.e. $\pi\in S_M$, the symmetric group of
permutations.
Furthermore, since $\pi(e_n)=j_n > e_n,\, n=1,\ldots,N$ and 
$\pi(g_m) = i_m < g_m, \, m=1,\ldots,M-N$, $\pi$ defined by (\ref{pi}) is a 
permutation with no fixed point, i.e. {\em derangements}. Yet another feature of $\pi$ is that 
it has exactly $N$ {\it excedances} defined as follows: an element $l \in [M]$ is an
{\em excedance} of $\pi$ if $\pi(l) > l$. The excedance set of $\pi$ in (\ref{pi})
is the set of pivot indices $\{e_1, e_2, \ldots, e_N\}$. 
Then we have the following 
characterization for the line-soliton solution of the
KP equation \cite{CK:09}.
\begin{Theorem}\label{Main}
\label{derangement}
Let $A$ be an $N\times M$, TNN, irreducible matrix which corresponds to
a point in the non-negative Grassmannian Gr$^+(N,M)$.
Then the $\tau$-function (\ref{B-C}) associated with this $A$-matrix 
generates an $(M-N,N)$-soliton solutions. The $M$ asymptotic line-solitons
associated with each of these solutions can be identified via a pairing map $\pi$ 
defined by \eqref{pi}. The map $\pi \in {S}_M$ is a derangement of the index 
set $[M]$ with $N$ excedances given by the pivot indices $\{e_1, e_2, \ldots, e_N\}$ 
of the $A$-matrix in RREF.
\end{Theorem}

\begin{figure}[h]
\begin{center}
\includegraphics[height=6.5cm]{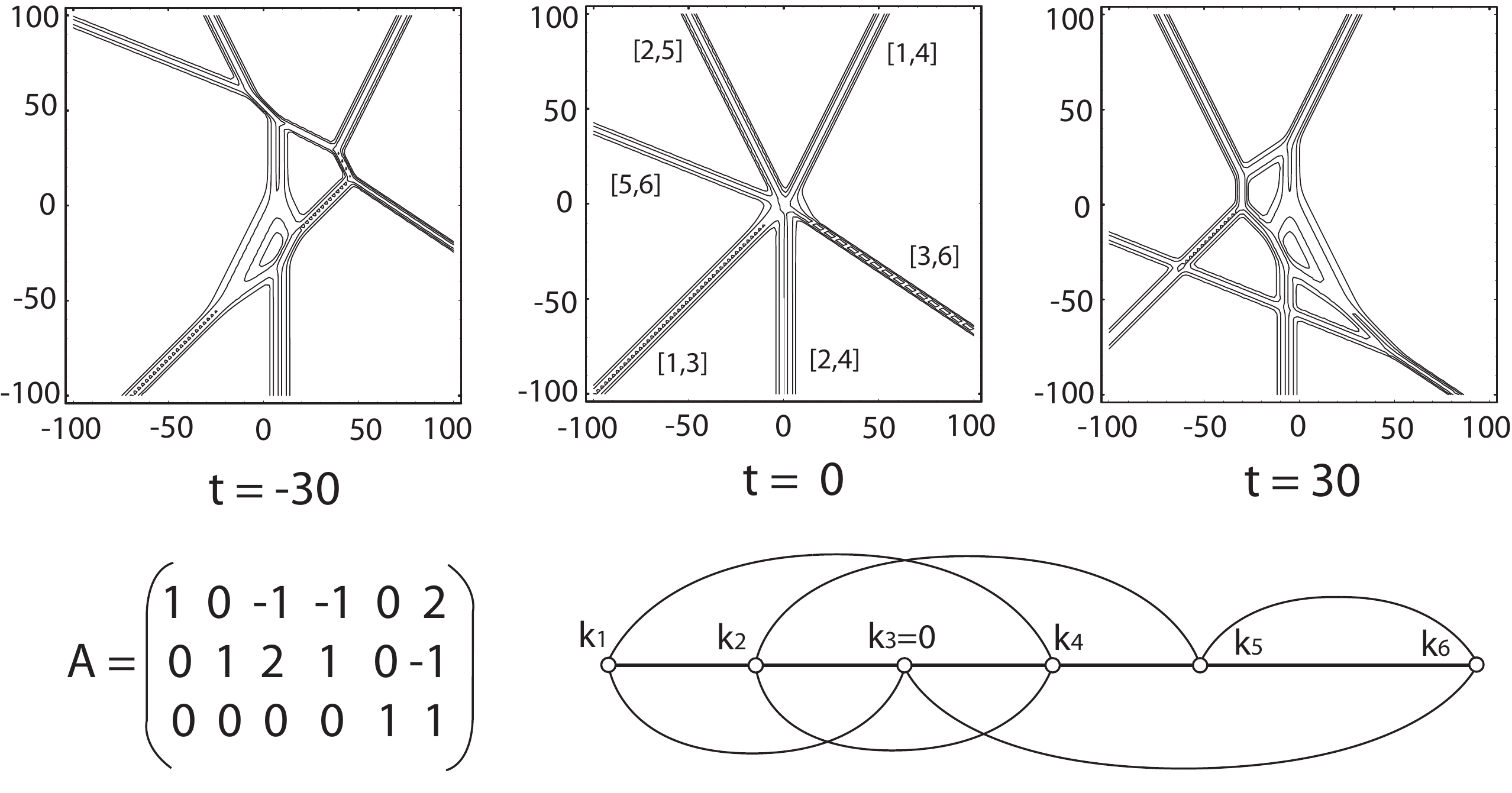}
\caption{An example of $(3,3)$-soliton solution. The permutation of this solution
is $\pi=(451263)$. The $k$-parameters are chosen as $({\kappa}_1,{\kappa}_2,\ldots,{\kappa}_6)=
(-1,-\frac{1}{2},0,\frac{1}{2},1,\frac{3}{2})$.  
(See \cite{KW:11} for a classification of those patterns.)\label{fig:33soliton}}
\end{center}
\end{figure}

Theorem \ref{Main} provides a unique parametrization of each TNN Grassmannian cell in terms of
the derangement of $S_M$. One should note that
Theorem \ref{Main} does not give us the indices $j_n$ and $i_m$ in the $[e_n, j_n]$ and $[i_m, g_m]$ 
line-solitons. 
The specific conditions that an index pair $[i,j]$ identifies an asymptotic 
line-soliton are obtained by identifying the dominant exponential in each domain in the $xy$-plane.
To visualize those asymptotic solitons, we define a {\it chord diagram}:
\begin{Definition}
A chord diagram associated to a derangement $\pi\in S_M$ is defined as follows:  Consider
a line segment with $M$ marked points by the numbers $\{1,\ldots,M\}$ in the increasing order from
the left.
\begin{itemize}
\item[(a)] If $i<\pi(i)$ (expedience), then draw a chord joining $i$ and $\pi(i)$ on the upper part of the line.
\item[(b)] If $j>\pi(j)$ (deficiency), then daw a chord joining $j$ and $\pi(j)$ on the lower part
of the line.
\end{itemize}
\end{Definition}
Then Proposition \ref{asymptote} and Theorem \ref{Main} imply that each chord diagram identifies the types of solitons appearing in the asymptotic regions
$|y|\gg 0$.  It is also useful to consider those marked points as the values of $k$-parameters, so that
for each chord joining $i$ and $j$, its length gives the amplitude and the sum of the joined points gives the inclination of the line-soliton of $[i,j]$-type. 
Figure \ref{fig:33soliton} illustrates the time evolution of an example of  $(3,3)$-soliton
solution. The chord diagram shows all asymptotic line-solitons for $y\to\pm\infty$.


\section{Shallow water waves: The Mach reflection}\label{sec:SWW2}
In this last section, we discuss a real application of the exact soliton solutions of the KP equation
described in the previous sections to the Mach reflection phenomena in shallow water.
Let us first discuss some exact solutions relevant to this phenomena.

\subsection{Exact soliton solutions from Gr$^+(2,4)$}
Recall that the $\tau$-function for Gr$^+(2,4)$ is given by
\begin{align}\label{eq:tau-function3}
\tau(x,y,t) =\sum_{1\le i<j\le 4}\Delta_{i,j}(A)E_{i,j}(x,y,t)
\end{align}
where $\Delta_{i,j}(A)\ge 0$ is the Pl\"ucker coordinates, the $2 \times 2$ minors consisting of   
$i$-th and $j$-th columns of the $2\times 4$ $A$-matrix, and $E_{i,j}={\rm Wr}(e^{\theta_i},e^{\theta_j})=(\kappa_j-\kappa_i)e^{\theta_i+\theta_j}$
 (note $E_{i,j}>0$ with the order $\kappa_i<\kappa_j$).
Proposition \ref{asymptote} shows that  $\tau$-function (\ref{eq:tau-function3}) 
generates a soliton solution which consists of at most
two line-solitons for both $y \to \pm\infty$. We consider the following two types:
 one consists of  two line-solitons of $[1,2]$ and $[3,4]$
for both $|Y|\gg 0$, which is called O-type soliton (``O'' stands for {\it original}, see \cite{K:04}); the other one consists of
$[1,3]$ and $[3,4]$ line-solitons for $Y \gg 0$ and 
$[1,2]$ and $[2,4]$ line-solitons for $Y \ll 0$. 
Theorem \ref{Main} indicates that
 this soliton can be referred as to $(3142)$-type, because those line-solitons
represent a permutation $\pi=\binom{1~2~3~4}{3~1~4~2}$.
\begin{figure}[h]
\begin{center}
 \includegraphics[width=37.0mm]{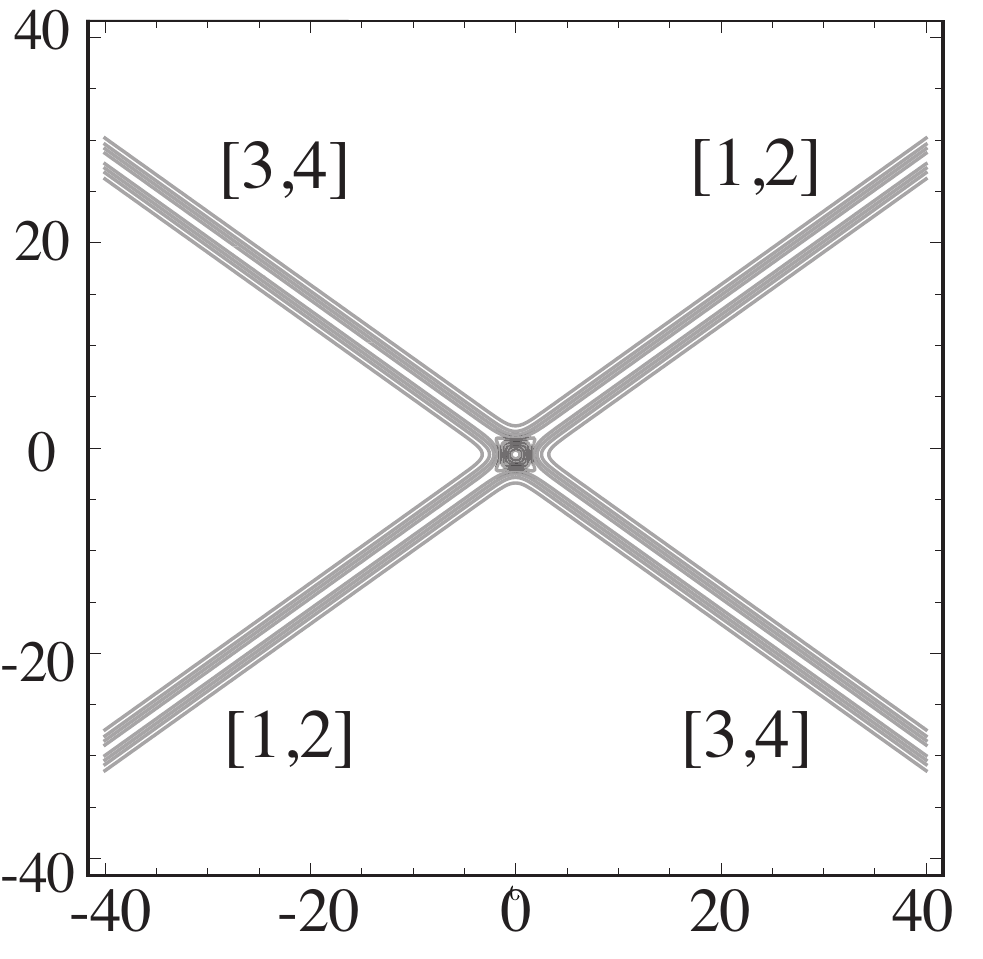}\hskip 2.8cm
 \includegraphics[width=37.0mm]{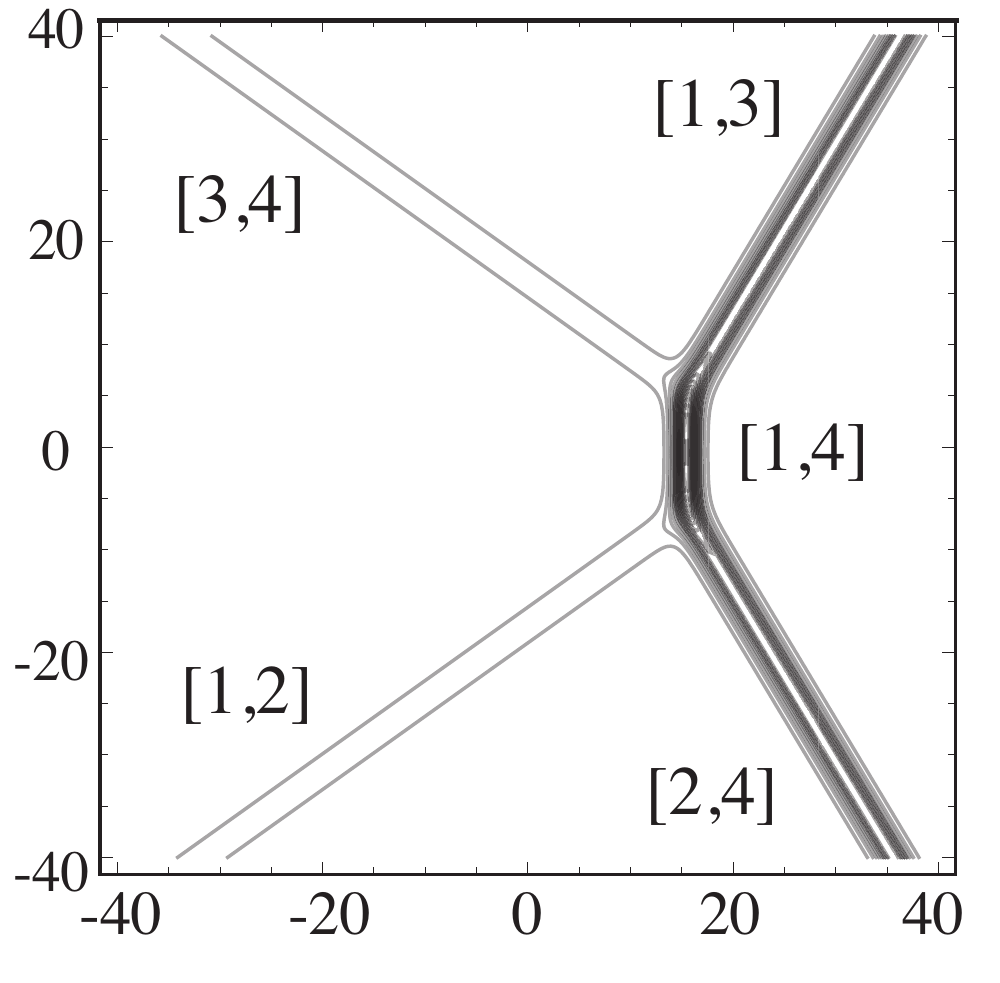}\\
 \bigskip
\hskip 0.3cm 
 \includegraphics[width=42.0mm]{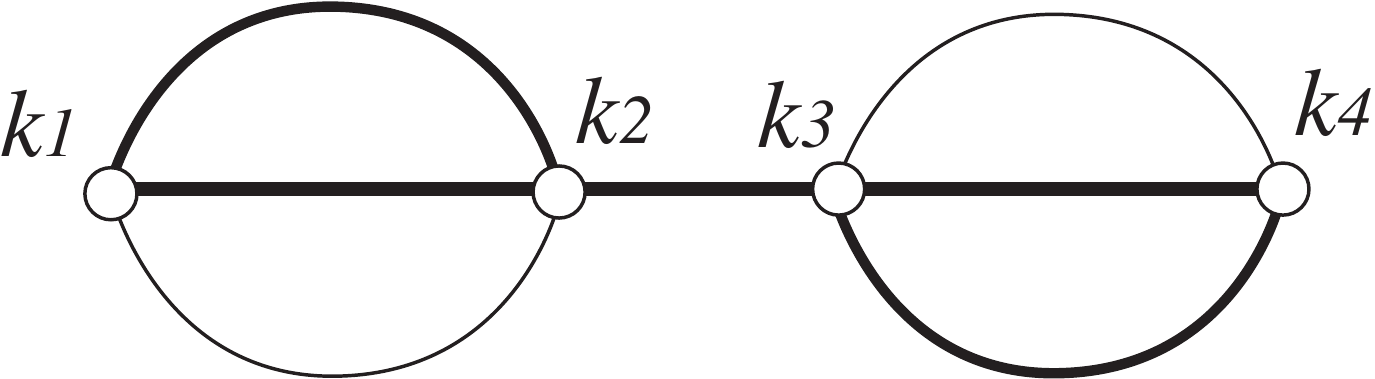}\hskip 2.2cm
\raisebox{-2.0mm}{ \includegraphics[width=42.0mm]{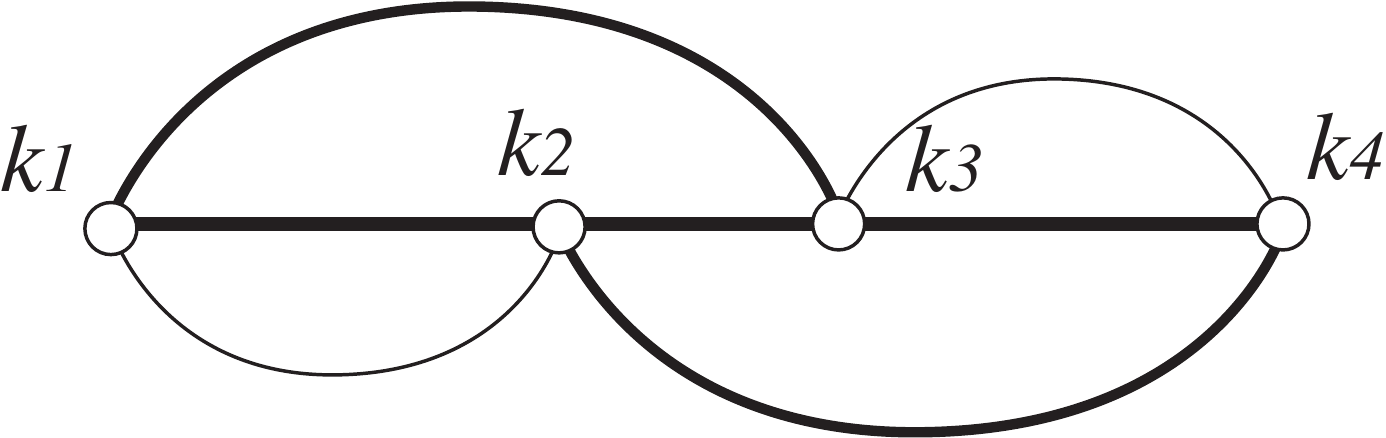}}
 \caption{Contour plots and the chord diagrams. Left: O-type 
 (or $(2143)$-type)
 solution. Right: (3142)-type solution.
 The length of $[1,4]$-soliton changes in $t$.}
 \label{fig:Fig3}
 \end{center}
\end{figure}
Figure \ref{fig:Fig3} illustrates the contour plots of O-type and (3142)-type solutions in the $xy$-plane, and the corresponding
{\it chord diagrams}. The upper chords represent the asymptotic solitons $[i,j]$ for $y\gg 0$, and the lower chords for the asymptotic solitons $[i,j]$ for $y
\ll 0$.  
The $A$-matrices for those solutions are respectively given by 
\begin{equation}
A_{\rm O}=
\begin{pmatrix}
1 & a & 0 & 0 \cr
0 & 0 & 1 & b\cr
\end{pmatrix}
, \qquad A_{(3142)}=
\begin{pmatrix}
1 & a & 0 & -c \cr
0 & 0 & 1 & b\cr
\end{pmatrix},  
\label{eq:O-3142matrix}
\end{equation} 
where $a, b, c > 0$ are constants determining the locations of the solitons (see \cite{CK:09}). Notice that the $\tau$-function  for the (3142)-type contains {\it five} exponential terms,
and the $\tau$-function for O-type with $c=0$ in (\ref{eq:O-3142matrix}) contains only {\it four} terms.

Let us fix the amplitudes and the angles of the solitons in the positive $x$ regions
for both O- and (3142)-types, so that those solutions are symmetric with respect to
the $x$-axis (see Figure \ref{fig:Fig3}):
\begin{equation}
\begin{array}{llll}
A_0&\equiv\left\{\begin{array}{l}A_{[1,2]}=A_{[3,4]} \qquad {\rm (O\,type)}\\[1.0ex]
A_{[1,3]}=A_{[2,4]}\qquad {\rm((3142)\,type)}\end{array}\right.\\
\\
\Psi_0&\equiv\left\{\begin{array}{l}-\Psi_{[1,2]}=\Psi_{[3,4]}>0\qquad {\rm (O\,type)}\\[1.0ex]
-\Psi_{[1,3]}=\Psi_{[2,4]}>0\qquad {\rm ((3142)\,type)}\end{array}\right.
\end{array}
\end{equation}
Then we express the $\kappa$-parameters  in terms of $A_0$ and $\tan\psi_0$
with $\kappa_1=-\kappa_4$ and $\kappa_2=-\kappa_3$ (due to the symmetry):
In the case of O-type, we have 
\begin{equation}
\kappa_1=-\frac{1}{2}\left(\tan\Psi_0+\sqrt{2A_0}\right), \quad 
\kappa_2=-\frac{1}{2}\left(\tan\Psi_0-\sqrt{2A_0}\right).
\label{eq:values of k (O)}
\end{equation}  
The ordering 
$\kappa_2<\kappa_3$ then implies $\tan\psi_0 > \sqrt{2A_0}$. 
On the other hand, for the (3142)-type, we have    
\begin{equation}
\kappa_1=-\frac{1}{2}\left(\tan\Psi_0+\sqrt{2A_0}\right), \quad 
\kappa_2=\frac{1}{2}\left(\tan\Psi_0-\sqrt{2A_0}\right).
\label{eq:values of k}
\end{equation} 
The ordering $\kappa_2<\kappa_3$  implies
$\tan\Psi_0<\sqrt{2A_0}$.
Thus, if all the solitons in the positive $x$-region have the same amplitude $A_0$
for both O- and (3142)-types, then an O-type solution arises when $\tan\Psi_0>\sqrt{2A_0}$,
and a (3142)-type when $\tan\Psi_0<\sqrt{2A_0}$. 
Then the limiting value at $\kappa_2=\kappa_3$ $(=0)$
 defines the critical angle $\Psi_c$,
\begin{equation}\label{critical}
\tan\Psi_c:=\sqrt{2A_0}.
\end{equation}
Note that at the critical angle, i.e. $\kappa_2=\kappa_3$, the $\tau$-function
has only {\it three} exponential terms, and this gives a ``Y"-shape resonant solution
as Miles noted \cite{M:77}.

One should note that for (3142)-type solution, the solitons
in the negative $x$-region are smaller than those in the positive region, i.e.
$A_{[3,4]}=A_{[1,2]}=\frac{1}{2}\tan^2\Psi_0 < A_0$, and the angles of those in the negative $x$-regions
do not depend on $\Psi_0$ and
$\Psi_{[3,4]}=-\Psi_{[1,2]}=\Psi_c$. 
Two sets of three solitons $\{[1,3],[1,4],[3,4]\}$ 
and $\{[2,4],[1,4],[1,2]\}$ are both in the soliton 
resonant state. These properties of the (3142)-type solution 
are the same as those of Miles's asymptotic solution 
for the Mach reflection in shallow water \cite{M:77}.   However one should note that the $[1,4]$-soliton corresponding to the
Mach stem becomes a soliton solution only in an asymptotic sense.
Then the exact solution of $(3142)$-type can provide an estimate of a propagation distance, at which the amplitude is sufficiently developed.

\subsection{The Mach reflection in shallow water}

\begin{figure}[h]
\centering
\includegraphics[scale=0.45]{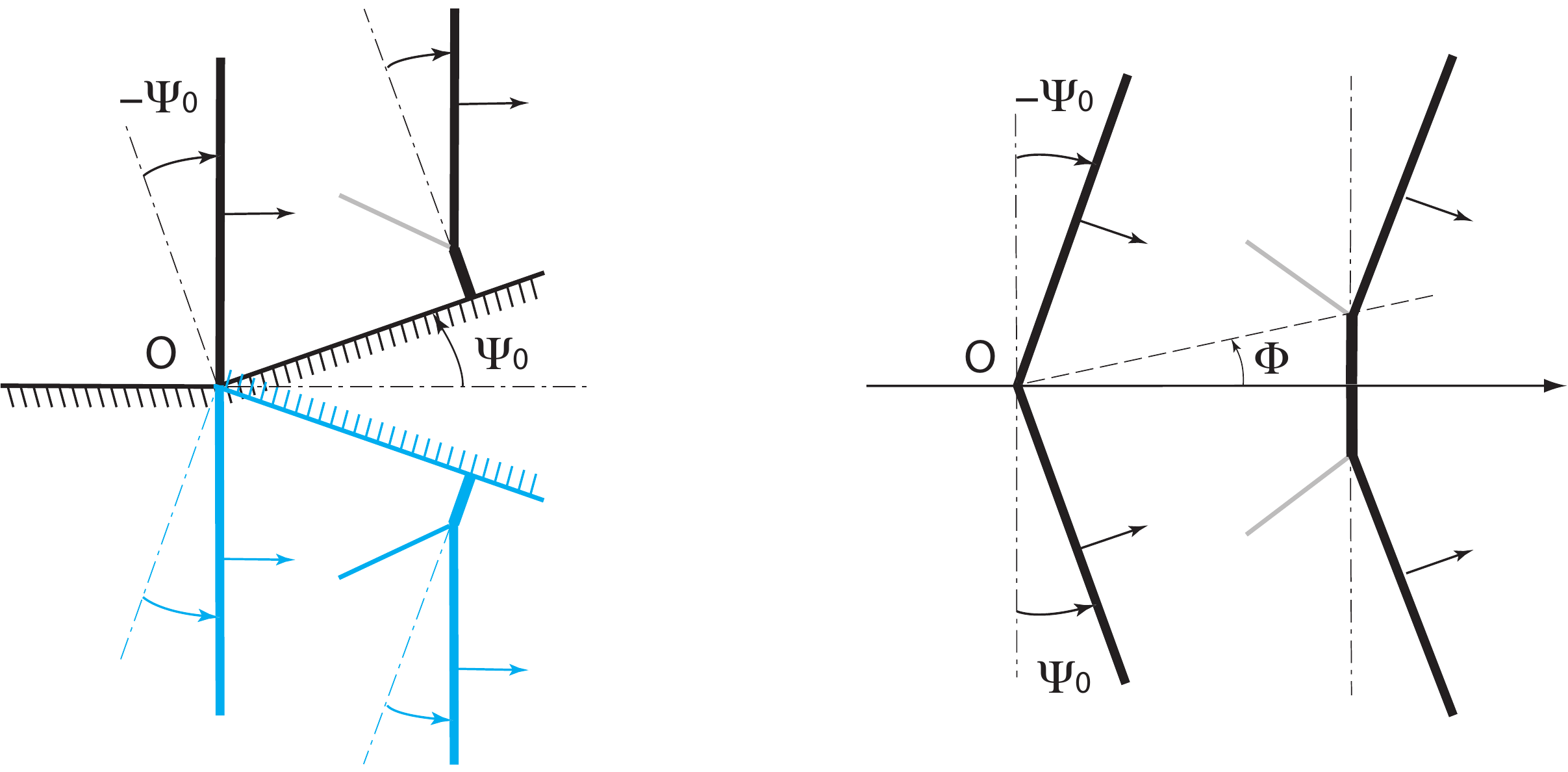} 
\caption{The Mach reflection. The left figure illustrates  an incidence wave propagating
parallel to the wall with the mirror image. The right figure is an equivalent system to the left one.   The resulting wave pattern shown here  is a $(3142)$-soliton solution.
The angle $\Phi$ becomes zero if the initial angle satisfies $\Psi_0\ge \Psi_c$, i.e.
no stem.
\label{fig:MR}}
\end{figure}

In \cite{M:77}, J. Miles considered an oblique interaction of two line-solitons  using O-type solutions.
He observed that resonance occurs at the critical angle $\Psi_c$, and when the initial oblique angle $\Psi_0$
is smaller than $\Psi_c$, the O-type solution becomes
singular. 
He also noticed a similarity between this resonant interaction and the Mach reflection
found in shock wave interaction (see for example \cite{CF:48,Wh:74}). This may be illustrated by the left figure of Figure \ref{fig:MR},
where an incidence wave shown by the vertical line is propagating to the right, and it hits a rigid wall
with the angle $\Psi_0$.
If the angle $\Psi_0$ is large, 
the reflected wave behind the incidence
wave has the same angle $\Psi_0$, i.e. a regular reflection occurs. However, if the angle is small,
then an intermediate wave called the Mach stem appears as illustrated in Figure \ref{fig:MR}.
The Mach stem, the incident wave and the reflected wave interact resonantly, and
those three waves form
a resonant triplet. 
The right panel in Figure \ref{fig:MR} illustrates the wave propagation which is  equivalent to that in the left panel, if one ignores the effect of viscosity on the wall (i.e. no boundary layer). 
These reflection patterns are well explained by the exact solutions of O- and (3142)-types
\cite{LYK:11,YLK:10}.  In the next section, we give a brief history of studies on the Mach reflection.

\subsection{Previous numerical results}
One of the most interesting things of the Mach reflection is that the KP theory predicts 
an extraordinary four-fold amplification of the stem wave at the critical angle \cite{M:77}.
The maximum amplitudes for those symmetric cases of O-type and (3142)-type occur
at $y=0$ (i.e. at the wall), and they are expressed by a parameter $k$, called the Miles 
parameter defined by (the original parameter was given by $k_M=\Psi_0/\sqrt{2A_0}$),
\begin{equation}\label{MilesK}
k=\frac{\tan\Psi_0}{\sqrt{2A_0}}.
\end{equation}
 Then from the KP theory, the amplification
factor $\alpha_0:=u_{\rm max}/A_0$ is given by \cite{M:77,CK:09}
\begin{equation}\label{alpha}
\alpha_0=\left\{\begin{array}{lll}
\displaystyle{(1+k)^2},\quad &{\rm for}\quad k<1, \\[1.0ex]
\displaystyle{\frac{4}{1+\sqrt{1-k^{-2}}}},\quad &{\rm for}\quad k>1.
\end{array}
\right.
\end{equation}
Several laboratory and numerical experiments tried to confirm the formula \eqref{alpha}, in particular,
 the four-fold amplification at the critical value $\kappa=1$ (see for example \cite{P:57, Me:80, F:80, T:93}).  In \cite{F:80}, Funakoshi made a numerical simulation
of the Mach reflection problem using the system of
equations equivalent to the Boussinesq-type system \eqref{eq:eta} and \eqref{eq:etat} up to the
first order.
He considered the initial wave with a small amplitude $a_i=0.05$, and  concluded that his results agree very well with the resonantly interacting solitary
wave solution predicted by Miles. However his results on the amplification parameter
$\alpha$ are slightly shifted to the lower values of the original Miles parameter $k_M=\Psi_0/\sqrt{3a_i}$.
Tanaka in \cite{T:93} then re-examined Funakoshi's results for higher amplitude incidence waves
with $a_i=0.3$ using the high-order spectral method.
He noted that  the effect of large amplitude tends to prevent the Mach reflection to occur, and
all the parameters such as the critical angle $\Psi_c$ are shifted toward the values corresponding
to the regular reflection (i.e. O-type). For example, he obtained the maximum amplification factor
$\alpha=2.897$ at $k_M=0.695$.   Because of the quasi-two dimensional approximation, i.e. $|\Psi_0|\ll 1$, Miles in his paper
  \cite{M:77} replaced $\tan\Psi_0$ by $\Psi_0$, and then in \cite{F:80,T:93}, the authors continued on to use this replacement. Then their computations
  with rather large values of $\Psi_0$ gave significant shifts of the parameter $\kappa$.
  In the next section, we re-evaluate their results using the new parameters obtained 
  from the normal form.

\subsection{Re-evaluation of the previous results using the corrected parameters obtained by the normal form}
We recall that the KP equation is derived under the assumptions of quasi-two dimensionality, weak dispersion and weak nonlinearity.  Then 
 for a physical application of the KP theory, one needs to include higher order corrections to those assumptions using  the normal form \eqref{N-KP2}.
  We then convert the observed amplitude in the numerical simulations
(or the experiments) to the corresponding KP amplitude via the formula \eqref{KPamp}.

\begin{figure}[t!]
\centering
\includegraphics[scale=0.38]{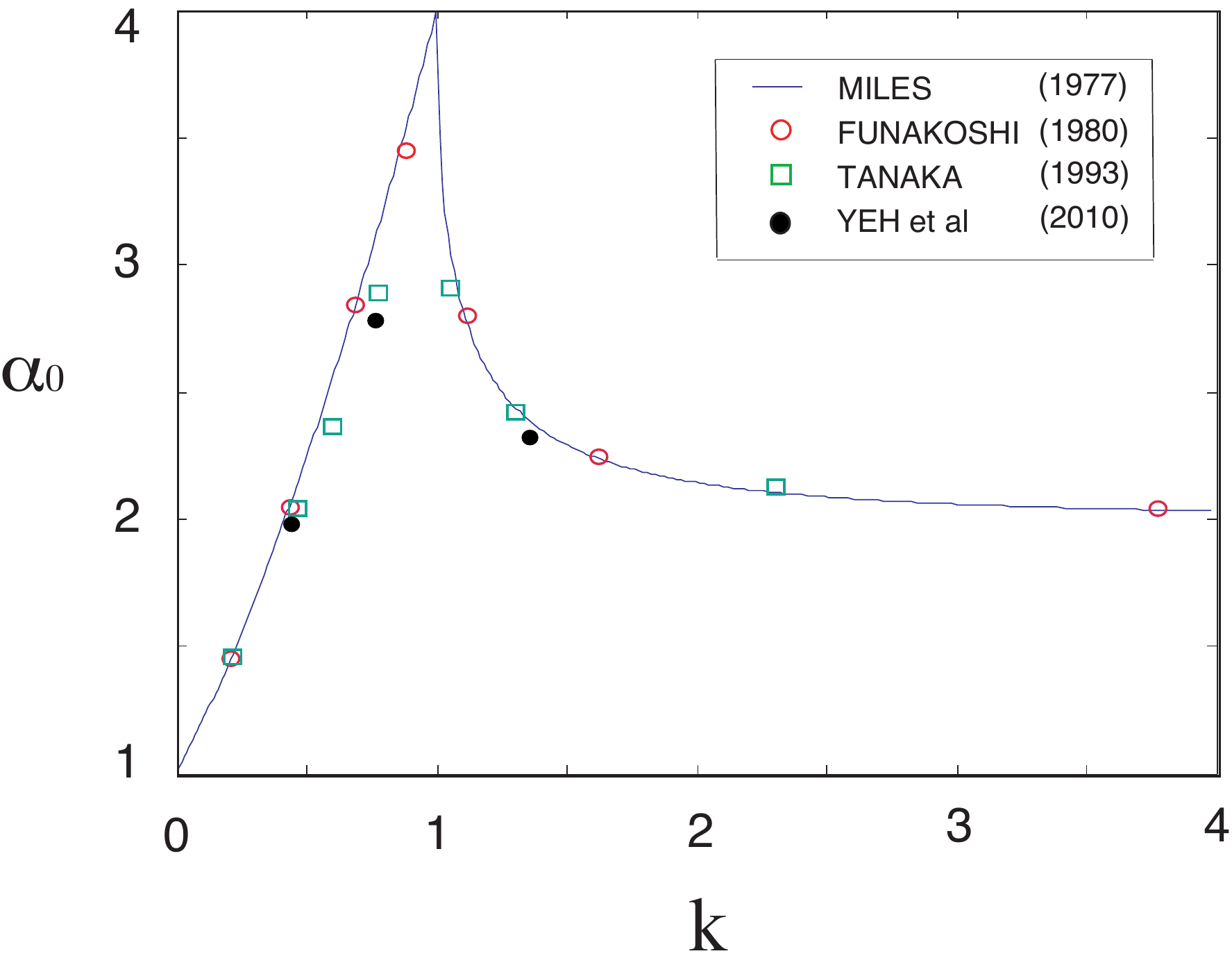}
\caption{Numerical results of the amplification factor $\alpha$ versus the parameter $\kappa$.
The circles show Funakoshi's result \cite{F:80}, the squares show Tanaka's result \cite{T:93}.
The black dots shows the experimental results by Yeh et al \cite{LYK:11}.\label{fig:FTY}}
\end{figure}

\begin{figure}[h]
\centering
\includegraphics[scale=0.75]{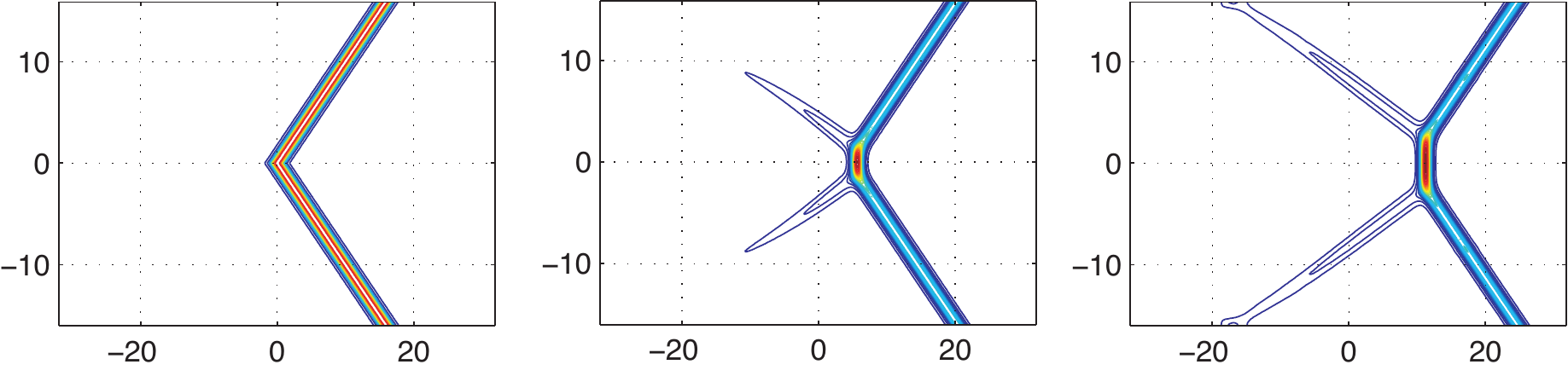}
\caption{Numerical simulation of with the V-shaped initial wave.
The solution converges to (3142)-type exact solution.
Notice a completion of a chord diagram.\label{fig:numerics}}
\end{figure}

Here we just summarize the results:  Let $a_i$ be the amplitude of the incident wave used in the numerical simulation (or the
experiment), and $\Psi_i=\Psi_0$ be the inclination of the incidence wave. Also let $\alpha$ be the amplification factor obtained from the numerical simulation (or the experiment).
Then the Miles parameter in terms of the observed amplitude is given by
\begin{equation}\label{newk}
k=\frac{\tan\Psi_0}{\sqrt{2A_0}}=\frac{\sqrt{1+\sqrt{1+5a_i}}\tan\Psi_0}{\sqrt{6a_i}\cos\Psi_0}.
\end{equation}
The corrected amplification factor (in the KP coordinate) is given by
\begin{equation}\label{newfactor}
\alpha_0=\left\{\begin{array}{lll}
\displaystyle{\frac{\alpha(1+\sqrt{1+5a_i})}{(1+\sqrt{1+5\alpha a_i})\cos^2\Psi_0}} \qquad&{\rm if}\quad k<1\\[2.0ex]
\alpha\qquad &{\rm if}\quad k>1
\end{array}\right.
\end{equation}
We then re-evaluate their results with the new formulae \eqref{newk} and \eqref{newfactor}, and the
results are shown in Figure \ref{fig:FTY}.  Since Funakoshi's simulations are based on small amplitude incidence waves,
his results agree quite well with the KP predictions.  Tanaka's results are also in good agreement
with the KP theory except for the cases near the critical angle (i.e. $k=1$), where the amplification
parameter $\alpha_0$ gets close to 3. This region clearly violates the assumption of the weak nonlinearity.   However, the original plots of Tanaka's
are significantly improved with those formulae \eqref{newk} and \eqref{newfactor}.
 The black dots in Figure \ref{fig:FTY} indicate the results of recent laboratory experiments done by  Yeh
 and his colleagues \cite{YLK:10,LYK:11}.
The laboratory experiments are performed  using 7.3 m long and 3.6 m wide  wave tank with a water depth of 6.0 cm.  In the talk, I will show the details of the experiments including some movies
and pictures of several real shallow water solitons.  I will also discuss a stability problem of solitons solutions, and present some numerical simulations which indicate a convergence of KP solutions
to certain exact soliton solutions similar to the case of the KdV equation \cite{KOT:09,KK:10}
(see Figure \ref{fig:numerics}).


\end{document}